\documentclass[12pt]{article}

\usepackage{amsmath}
\usepackage{epsf}
\begin{document}
\sloppypar

 {\it Accepted for publication in Astronomy
Letters, v.  28, N 2, 2002.}
\vspace{2cm}
\bigskip
 
\large
\centerline{\bf Nodal and Periastron Precession of Inclined Orbits }
\centerline{\bf in the Field of a Rapidly Rotating Neutron Star .}
\vspace{15mm}
 \normalsize
\centerline{ Nail Sibgatullin$^{1,2}$ }
\vspace{2mm}
\noindent
$^1${\it Moscow State University, Vorob'evy gory, Moscow, 119899 Russia }\\
$^2$ {\it Max-Planck-Institut f\"ur Astrophysik, Karl-Schwarzschild-Str. 1,
85740 Garching bei \\ \noindent Munchen, Germany}\\

$^*$ {e-mail:sibgat@mech.math.msu.su}
 
\vspace{7mm}
\def\f{\frac}
\def\p{\partial}
\begin{abstract}
We derive a formula for the nodal precession frequency and the Keplerian period
of a particle at an arbitrarily inclined orbit (with a minimum latitudinal angle reached
at the orbit) in the post-Newtonian approximation in the external field of an oblate
rotating neutron star (NS). We also derive formulas for the nodal precession and
periastron rotation frequencies of slightly inclined low-eccentricity orbits in the
field of a rapidly rotating NS  in the form of asymptotic expansions whose first
terms are given by the Okazaki--Kato formulas. The NS gravitational field is described
by the exact solution of the Einstein equation that includes the NS quadrupole moment
induced by rapid rotation. Convenient asymptotic formulas are given for the metric
coefficients of the corresponding space-time in the form of Kerr metric perturbations in
Boyer--Lindquist coordinates.

\end{abstract}

\section*{INTRODUCTION}

The X-ray flux from low-mass X-ray binaries (LMXBs) commonly exhibits two peaks in the
power spectrum at frequencies $\sim1$~kHz and one peak in the range 10--100~Hz
(Van~der~Klis 2000). Stella and Vietri (1998) proposed to interpret the frequency
difference between the kHz peaks as the periastron rotation frequency of
low-eccentricity orbits: the observed decrease in this difference for some LMXBs was
explained by a radiating clump approaching the marginally stable orbit at which the
periastron rotation frequency becomes zero. The low-frequency peak was interpreted as
the precession frequency of circular Keplerian orbits inclined to the equator (Merloni
\emph{et al.} 1998; Morsink and Stella 1999; Stella and Vietri 1999; Psaltis \emph{et
al.} 1999). The formulas of Okazaki \emph{et al.} (1987) derived in terms of the Kerr
solution for orbits slightly inclined to the equatorial plane are commonly used to
compare models with observations. The importance of the precession of inclined orbits in
the fields of rotating black holes in interpreting the quasi-periodic oscillations
(QPOs) of the X-ray flux from  pulsars and black-hole candidates was first pointed out
by Cui \emph{et al.} (1998). Inclined orbits in Kerr and Kerr--Newman fields were first
considered by Wilkins (1972) (uncharged case) and Johnston and Ruffini (1974) (for a
charged rotating black hole). Van~Kerkwijk \emph{et al.} (1998) explained the puzzling
spindown and spinup of some X-ray pulsars by the fact that the accretion-disk tilt to
the equatorial plane in inner regions can become larger than 90$^{\circ}$! Previously
(Sibgatullin 2001), we derived analytic expressions for the nodal precession and
periastron rotation frequencies for orbits \emph{arbitrarily inclined} to the equatorial
plane\footnote{To be more precise, with an arbitrary minimum latitudinal angle, which
can be reached in a bound trajectory.}. The nodal precession of the \emph{marginally
stable orbits} is described as a function of the corresponding Keplerian frequency for
various inclinations of these orbits. For a Keplerian frequency of 1200~Hz and a NS mass
of~2.2~$M_{\odot}$, the nodal frequency in the marginally stable orbit was shown to
change from 41~Hz to 123~Hz as the inclination to the equator changes from~0
to~90$^{\circ}$, i.e., by a factor of~3!

However, the results for black holes need to be significantly corrected for neutron
stars, because a rapidly rotating NS becomes oblate and a quadrupole moment appears. For
orbits with low inclinations to the equatorial plane, Markovic (2000) derived formulas
for the nodal and periastron precession frequencies in the post-Newtonian approximation
by taking into account the NS quadrupole moment. Morsink and Stella (1999) and Vietri
and Morsink (1999) \emph{numerically} calculated $\nu_r$ and $\nu_{\mathrm{nod}}$ as
functions of $\nu_{\phi}$ for low inclinations and low eccentricities for various NS
equations of state.

Here, our goal is to derive \emph{analytic} formulas for $\nu_r$ and $\nu_{\mathrm{nod}}$
for \emph{low} inclinations and low eccentricities in the form of asymptotic expansions,
which transform to the formulas of Okazaki \emph{et al.} (1987) and Kato (1990) at a
zero quadrupole moment. The exact quadrupole solution extracted from the more general
solution found by Man'ko \emph{et al.} (1994) formed a basis for our study. This
solution was reduced to the simplest form in Weyl coordinates. For clarity, the exact
quadrupole solution is also given asymptotically as a Kerr metric perturbation in
Boyer--Lindquist coordinates.

Sibgatullin and Sunyaev (1998, 2000a, 2000b) (below referred to as SS~98, SS~00a, and
SS~00b) provided formulas for the luminosity and spindown rate for various NS equations
of state. These authors proposed a method of analytically constructing the quadrupole
moment $b$ as a function of the Kerr parameter $j$ and NS rest mass $m$, $b=b(j,m)$,
based on the main NS thermodynamic function --- its gravitational mass $M$ as a function
of the Kerr parameter and rest mass $M=M(j, m)$. In this case, the space-time geometry
for the exact quadrupole solution plays a crucial role in finding the marginally stable
orbit. The dependence $b(j, m)$ was constructed using numerical data from Cook \emph{et
al.} (1994) and the numerical code by Stergioulas (1998) for the marginally stable
orbit. Laarakkers and Poisson (1998) found a parabolic dependence of the quadrupole
moment on the Kerr parameter at $j << 1$ by using direct calculations from the formula
of Ryan (1995b, 1997) for several equations of state. The physical parameters of the
marginally stable orbit were studied by Shibata and M.~Sasaki (1998) using expansions at
large radii.

An important outstanding question (which is not considered here) is the passage from a
viscous accretion disk (Shakura and Sunyaev 1973, 1976) to free particles near the
marginally stable orbit. Another complicated question (which is not considered here
either) is the behavior of bound trajectories at the marginally stable orbit for their
finite inclination to the equatorial plane.

\section{THE EXTERNAL GRAVITATIONAL FIELDS OF ROTATING NEUTRON STARS}

The efficiency of the exact quadrupole solution in describing the external fields of
rapidly rotating neutron stars with masses larger than 1~$M_{\odot}$ was demonstrated in
SS~98, SS~00a, and SS~00b. In contrast to the Kerr solution, this solution contains an
additional constant, which has the meaning of intrinsic (non-Kerr) NS quadrupole moment,
and is symmetric relative to the equatorial plane. The solution under discussion is
contained as a special case in the five-parameter (mass, angular momentum, quadrupole
moment, electric charge, and magnetic dipole ) solutions obtained by Man'ko \emph{et al.}
(1994) using the method developed in our book (Singatullin 1991)\footnote{For the
relationship of this method to the soliton solutions of Kramer and Neugebauer (1980),
see Ernst (1994) and Man'ko and Ruiz (1998).}. However, the expressions for the metric
coefficients and the 4-potential of the electromagnetic field are cumbersome. To solve
problems with disk accretion in the equatorial plane, we used simple expressions for the
metric of the quadrupole solution in the equatorial plane (SS~98, SS~00a). Since the
importance of the quadrupole solution in describing the fields of rotating NS
gravitational fields is beyond question, we give here the corresponding exact solution
of the Einstein equations in the entire space \footnote{The metric of the exact
solutions for the Einstein equations with a finite set of multipole moments is given in
general form in SS~00 . The method of constructing exact asymptotic flat solutions for
the Einstein-Maxwell set of  equations with data specified on the symmetry axis was developed in our book
(Sibgatullin 1991). The entire problem reduces to solving the only homogeneous singular
equation with the Cauchy kernel on a segment with an additional normalization
condition.}. Thus, the metric coefficients in the square of the interval in
vacuum\footnote{in Papapetru form.},

\begin{gather}
ds^2=-f(dt-\omega
d\phi)^2+\frac{\rho^2}{f}d\phi^2+{}\\
{}+\frac{\exp 2\gamma}{f}(d\rho^2+dz^2)\nonumber
\end{gather}

in the exact quadrupole solution are\footnote{Here, we use a system of units with
$M=c=G=1$.}

\begin{gather}
f=\mathrm{Re}\left(\frac{A-B}{A+B}\right),\\
 f\omega=2\mathrm{Re}\left(\frac{j(A-B)+iC}{A+B}\right),\nonumber\\
e^{e\gamma}=\frac{A\bar{A}-B\bar{B}}{16k_{+}^4k_{-}^4R_{+}R_{-}r_{+}r_{-}}.\nonumber
\end{gather}

In formulas~(2), we use the following notation

\begin{gather}
A=k_{-}^2(R_{-}+r_{+})(R_{+}+r_{-})-{}\\
{}-k_{+}^2(R_{-}-r_{-})(R_{+}-r_{+}),\nonumber\\
B=k_{+}k_{-}((k_{+}+k_{-})(r_{+}+r_{-})+{}\nonumber\\
{}+(k_{-}-k_{+})(R_{-}+R_{+}),\nonumber\\
C=zB+k_{+}k_{-}(k_{-}(R_{+}r_{+}-R_{-}r_{-})+{}\nonumber\\
{}+k_+(R_{+}r_{-}-R_{-}r_{+})+2b(R_{+}-R_{-}+r_{-}-r_{+})),\nonumber\\
R_{\pm}\hspace{-3pt}=\hspace{-3pt}\sqrt{\rho^2\hspace{-2pt}+\hspace{-2pt}
(z\pm(k_{+}\hspace{-2pt}+\hspace{-2pt}
k_{-})/2)^2}(\sqrt{1-j^2}\pm ij),\\
r_{\pm}\hspace{-3pt}=\hspace{-3pt}\sqrt{\rho^2+
(z\pm(k_{+}-k_{-})/2)^2}(\sqrt{1-j^2}\pm ij),\nonumber\\
k_{-}=\sqrt{1-j^2-4b}, \quad k_{+}=\sqrt{1-j^2}.\nonumber
\end{gather}
The metric coefficient $\omega$ becomes zero on the symmetry axis at
$|z|>(k_{+}+k_{-})/2$. We assume that $\omega$ on the symmetry axis becomes zero
everywhere outside the rotating NS (the condition for the absence of conical points).
Solution~(2)--(4) was obtained for the Ernst potential on the symmetry axis:

\begin{gather*}
\mathbf{E}=\frac{A-B}{A+B}|_{\rho=0}=\frac{z^2+(ij-1)z+b^2}{z^2+(ij+1)z+b^2}.
\end{gather*}
Note that $(j(A-B)+iC)/(A+B)=i$ on the symmetry axis.

In the special case where $b=(1-j^2)/4$, the constant $k_{-}$ becomes zero. In this
case, it is convenient to pass to the coordinates $\rho=\sin
\theta\sqrt{(r\hspace{-2pt}-\hspace{-2pt}1)^2\hspace{-2pt}-\hspace{-2pt}(1\hspace{-2pt}-
\hspace{-2pt}j^2)/4}$, $z=(r-1)\cos\theta$. Passing to the limit $k_{-}\rightarrow0$,
the metric coefficients in the square of the interval

\begin{gather}
ds^2=-f(dt-\omega d\phi)^2+\frac{\rho^2}{f}d\phi^2+{}\\
{}+
\frac{g}{f}\left(\frac{dr^2}{(r-1)^2-(1-j^2)/4}+d\theta^2\right)\nonumber
\end{gather}
may be represented as
\begin{gather}
f=\mathrm{Re}\left(\frac{a}{b}\right),\quad
f\omega=2\mathrm{Re}\left(\frac{c}{b}\right),\\
g=\frac{\mathrm{Re}(ab^*)}{4(4(r-1)^2-(1-j^2)m^2)^3.\nonumber}
\end{gather}
Here, the asterisk denotes a complex conjugate and the following notation is used with
$m\equiv\cos\theta$:

\begin{gather*}
a=-27-5j^2+18ijm-2ij^3m-6ijm^3\\
{}+6ij^3m^3-j^2m^4+j^4m^4+108r+4j^2r-{}\\
{}-48ijmr+4ijm^3r- 4ij^3m^3r-144r^2+{}\\
{}+48ijmr^2+80r^3-16ijmr^3-16r^4,\\
b=-3+3j^2+14ijm+2ij^3m-2ijm^3+{}\\
{}+2ijm^3-j^2m^4+j^4m^4+10r-4j^2r-{}\\
{}-48ijmr+ 4ijm^3r-4ij^3m^3r-48r^2+{}\\
{}+48ijmr^2+48r^3-16ijmr^3-16r^4,\\
c=ij^4m(-1+2m^2)-im(-3+2r)(-1+2r)^3-{}\\
{}-2ij^2m(-2+m^2+2r)+
j(-13+m^4(1-2r)+{}\\
{}+\hspace{-2pt}46r\hspace{-1pt}-48r^2\hspace{-1pt}+16r^3)\hspace{-1pt}
+j^3(-\hspace{-1pt}3+2r\hspace{-1pt}+m^4(-\hspace{-1pt}1+2r)).
\end{gather*}
Let us again turn back to the general exact quadrupole solution (2)--(4) and pass to
Boyer--Lindquist coordinates $\rho=\sin\theta\times\linebreak
\times\sqrt{(r-1)^2-(1-j^2)}$, $z=(r-1)\cos\theta$.

Let us represent the square of the interval as
\begin{gather}
ds^2=-fdt^2+2f\omega
dtd\phi+{}\\
{}+\sin^2\theta\left(\frac{r^2-2r+j^2}{f}-f\omega^2\right)d\phi^2+{}\nonumber\\
{}+\frac{\Gamma}{f}
\left(\frac{dr^2}{r^2-2r+j^2}+d\theta^2\right),\nonumber
\end{gather}
where
\begin{gather}
f=\frac{F}{r^2+j^2\cos^2\theta},\quad\omega=\frac{W}{r^2+j^2\cos^2\theta}.
\end{gather}
We will seek the functions $F$, $\Gamma$, and $W$ using the exact solution (2)--(4) in the
form of series in inverse powers of the radius. At~$b=0$, the series for $F$, $\Gamma$, and
$W$ break off on the second or third term, giving the exact Kerr solution

\begin{gather*}
F_k=\Gamma_k=r^2-2r+j^2\cos^2\theta,\quad W_k=-2jr\sin^2\theta.
\end{gather*}
At~$b\neq 0$, the series for $F$, $G$, and $W$ contain an infinite number of terms and
diverge on the event horizon. This fact is related to the so-called \emph{no hair
theorem}\footnote{According to Wheeler's figurative expression, ``a black hole has no
hair ''.} (see Misner \emph{et al.} (1973), Novikov and Frolov (1986), Sibgatullin (1984) for a more detailed
discussion of the black-hole theory). A collapsing star loses its magnetic and
quadrupole moment and its field asymptotically approaches the field of a rotating black
hole (Ginzburg and Ozernoi 1964; Doroshkevich \emph{et al.} 1965).

Cumbersome transformations (in these formulas, $m\equiv\cos\theta$) yielded

\begin{gather}
F=r^2-2r+j^2m^2+b\left(\frac{3m^2-1}{r^2}(r+1)\right.\\
\left.\frac{1}{r^3}\left(\frac{3}{4}(1+j^2+b)-
\frac{1}{2}(3+25j^2+15b)m^2\right.\right.\nonumber\\
{}\left.\left.+\frac{1}{4}(-5+63j^2+35b)m^4\right)+
\mathrm{o}\left[\frac{1}{r^4}\right]\right);\nonumber\\
\Gamma=r^2-2r+j^2m^2+b(1-m^2)\nonumber\\
{}\times\left(\frac{3}{2r^3}(5m^2-1)(r+2)+\frac{1}{4r^4}(-17-j^2\nonumber\right.\\
\left.
{}+8b+2(35+10j^2+56b)m^2+(35-{}\right.\nonumber\\
{}\left.-35j^2-180b)m^4)+\mathrm{o}\left[\frac{1}{r^5}\right]\right);\nonumber\\
W=2j(1-m^2)\left(-r+b\left(\frac{1}{4r^3}(5m^2-1)\nonumber\times\right.\right.\\\left.\left.\hspace{-2pt}(4r\hspace{-1pt}+\hspace{-1pt}7)\hspace{-1pt}+
\hspace{-1pt}\frac{1}{8r^4} (\hspace{-1pt}-\hspace{-1pt}24-b+2
(54+20j^2+63b)m^2\hspace{-2pt}\nonumber\right.\right.\\
\left.\left.+(12-72j^2-189b)m^4) +\mathrm{o}\left[\frac{1}{r^5}\right]\right)\right).\nonumber
\end{gather}

\section{PERIASTRON AND NODAL PRECESSION OF ORBITS WITH LOW INCLINATIONS \\ \protect TO THE EQUATORIAL PLANE
\\ \protect AND WITH LOW ECCENTRICITIES}

The eikonal equation $g^{ij}S_{,i}S_{,j}=-1$ in stationary spaces with axial symmetry
and with the square of the interval (1) have solutions of the form
$S=-Et+L\phi+\tilde{S}(\rho,z)$, where $\tilde{S}$ satisfies the equation

\begin{gather}
(\tilde{S}_{,\rho}^2+\tilde{S}_{,z}^2)fe^{-2\gamma}=V(\rho,z)\equiv
\frac{E^2}{f}-\\
{}-\frac{f}{\rho^2}(L-\omega E)^2-1.\nonumber
\end{gather}
Consider nearly circular orbits in the equatorial plane. For the latter,
$V=V_{,z}=V_{,\rho}=0$ at $z=0$. Consequently, expanding the right-hand part of Eq.~(10)
for perturbations in a Taylor series to within quadratic terms, we have

\begin{gather}
(\tilde{S}_{,\rho}^2+\tilde{S}_{,z}^2)fe^{-2\gamma}=\frac{1}{2}(V_{,\rho\rho}(\Delta\rho)^2
+V_{,zz}z^2)+\mathrm{const}.
\end{gather}
The constant in Eq.~(11) is related to the perturbation of $V$ when the constants $E$
and $L$ are perturbed for nonequatorial, noncircular orbits. Equation~(11) can be solved
by the separation of variables, $\tilde{S}=S_1(\Delta\rho)+S_2(z)$. Consider the
Hamiltonian system associated with Eq.~(11). We will seek $\Delta\rho$ and $z$ in the
form $\Delta\rho=\epsilon_1\sin\xi$, $z=\epsilon_2\sin\zeta$. We then obtain

\begin{gather}
\frac{d\zeta}{ds}=\sqrt{-fe^{-2\gamma}V_{zz}/2},\quad
\frac{d\xi}{ds}=\sqrt{-fe^{-2\gamma}V_{\rho\rho}/2},\\
\frac{d\phi}{ds}=\frac{f}{\rho^2}(L-\omega E),\quad
\frac{dt}{ds}=\frac{E}{f}+\frac{f\omega}{\rho^2}(L-\omega
E).\nonumber
\end{gather}
Note that the right-hand parts of Eqs.~(12) are constant.

Let us now make use of the expression for the energy and angular momentum of particles
in equatorial circular orbits (SS~98, SS~00a):

\begin{gather}
E=\frac{\sqrt{f}}{\sqrt{1-f^2p^2/r}},\quad L-\omega E=pE,
\end{gather}
\begin{gather*}
p\equiv\rho^2(-\lambda+\sqrt{\lambda^2+\mu-\mu^2r})/n,\\
\lambda\equiv f\dot{\omega}, \mu\equiv\dot{f}/f, n\equiv f-r\dot{f}.
\end{gather*}
Here, the dot denotes a derivative with respect to~$\rho^2$.

For the angular velocity of a particle in a Keplerian equatorial circular orbit, the
following formula can be derived from Eqs.~(12):

\begin{gather*}
\omega_{\phi}=\frac{p}{(\rho^2/f^2+\omega p)}.
\end{gather*}
Below, it is convenient to introduce a new quantity instead of~$p$:
$\tau\equiv\rho^2/(fp)$. The formula for the angular velocity can then be rewritten as

\begin{gather}
\nu_{\phi}=\frac{1}{T},\quad T/(2\pi)=\omega+\tau/f,\\
\tau\equiv(\lambda+\sqrt{\lambda^2+\mu-\mu^2r})/\mu.\nonumber
\end{gather}
For the rotation frequency from periastron to periastron, $2\pi\nu_r=d\xi/dt$, we obtain
from Eqs.~(12)

\begin{gather}
\nu_r^2=\nu_{\phi}^2M,
\end{gather}
where we use the notation

\begin{gather*}
M\equiv 2e^{-2\gamma}\left(-gf\left(\frac{1}{f}\right)_{,qq}\tau^2+
\frac{q^3}{f}\left(\frac{f}{q}\right)_{,qq}-\right.\\
{}-4q^2\left(\frac{f}{q}\right)_{,q}\tau\omega_{,q}-2q\tau
f\omega_{qq}+2\tau^2f^2(\omega_{,q})^2\Bigg)|_{z=0},
\end{gather*}
$q\equiv\rho^2$. In exactly the same way, for the rotation frequency of the maximum rise
in~$z$, $2\pi\nu_{\theta}=d\zeta/dt$, we obtain from Eqs.~(12) using the explicit
expression~(10) for~$V$

\begin{gather}
\nu_{\theta}^2=\nu_{\phi}^2N,\\
N=\frac{e^{-2\gamma}}{f}(f_{,zz}(\rho^2+\tau^2)-
2f\omega_{zz}\tau)|_{z=0}.\nonumber
\end{gather}
For the periastron and nodal precession frequencies, we have

\begin{gather}
\nu_{\mathrm{per}}=\nu_{\phi}(1-\sqrt{M}),\quad
\nu_{\mathrm{nod}}=\nu_{\phi}(1-\sqrt{N}).
\end{gather}
We will seek the functions $T(r)$, $N(r)$, and $M(r)$ in coordinates~(7) in the form of
expansions in terms of inverse powers of~$r$. At~$b=0$, the series break off to give the
exact formulas of Okazaki \emph{et al.} (1987) and Kato (1990) for the Kerr metric

\begin{gather*}
M=1-\frac{6}{r}+\frac{8j}{r^3/2}-\frac{3j^2}{r^2}, \quad
N=1-\frac{4j}{r^{3/2}}+\frac{3j^2}{r^2}, \\ T/(2\pi)=j+r^{3/2}.
\end{gather*}
Let $b\neq 0$; using coordinates~(7), we obtain

\begin{gather}
\hspace{-5pt}M=1-\frac{6}{r}+\frac{8j}{r^{3/2}}-\frac{3(j^2+b)}{r^2}+
3b\left(-\frac{5}{r^3}\right.\\
{}+\frac{8j}{r^{7/2}}-\frac{1}{2r^4}(16+7j^2+2b)+
\frac{12j}{r^{9/2}}-\nonumber\\
{}\left.-\frac{1}{4r^5}(51+3j^2+31b)+\mathrm{o}\left[\frac{1}{r^{11/2}}\right]\right);\nonumber\\
N=1-\frac{4j}{r^{3/2}}+\frac{3(j^2+b)}{r^2}+3b
\left(\frac{2}{r^3}-\right.\\
{}\left.-\frac{5j}{r^{7/2}}+\frac{1}{2r^4}
(8+7j^2+2b)-\frac{9j}{r^{9/2}}+\mathrm{o}\left[\frac{1}{r^{5}}\right]\right);\nonumber\\
T/(2\pi)=r^{3/2}+j+b\left(-\frac{3}{4r^{1/2}}-\frac{5}{4r^{3/2}}\right.\\
{}+\frac{3j}{2r^2}-\frac{3}{32r^{5/2}}
(10+10j^2+b)+\frac{5j}{r^3}-\nonumber\\
{}-\left.\frac{3}{16r^{7/2}}(29+13j^2+4b)+
\mathrm{o}\left[\frac{1}{r^{4}}\right]\right).\nonumber
\end{gather}
Using Eqs.~(19) and~(20) for the nodal velocity of slightly inclined orbits, we derive a
Taylor series in inverse powers of the radius:

\begin{gather}
\hspace{-8pt}2\pi\nu_{\mathrm{nod}}=\hspace{-2pt}\frac{2j}{r^3}-\hspace{-2pt}\frac{3(b+j^2)}{2r^{7/2}}
-\hspace{-2pt}\frac{3b}{r^{9/2}}+\hspace{-2pt}\frac{3j}{2r^5}
(-j^2\hspace{-2pt}+5b)\hspace{-2pt}-\\
{}- \frac{3}{8r^{11/2}}(16b-3j^4+11j^2b+4b^2)
+\mathrm{o}\left[\frac{1}{r^{6}}\right].\nonumber
\end{gather}
At~$b=0$, we obtain a Taylor expansion of the formula by Okazaki \emph{et al.} (1987)
and Kato (1990) from Eq.~(21).

For orbits with low inclinations to the equatorial plane, Markovic (2000) derived
formulas for the periastron and nodal precession frequencies in the \emph{post-Newtonian
approximation}. His formulas for the nodal precession frequency is equivalent to
Eq.~(21) if the first two terms are retained in it. The latter are the sum of the
Newtonian precession of a slightly inclined orbit in a gravitational field with a
quadrupole moment and the precession of Lense and Thirring (1918).

Thus, for orbits with low inclinations to the equatorial plane and with low
eccentricities, allowance for NS oblateness in Eqs.~(17) for the nodal and periastron
precession of orbits with rapid rotation gives a \textbf{large contribution for the hard
equations of state}. The method for constructing the dependence of quadrupole moment on
the NS rest mass and Kerr parameter is described in SS~00a; for the equations of state A
and FPS, specific functions $b(j,m)$ were constructed at rest masses larger than the
solar mass and smaller than the critical mass according to the static stability
criterion. Laarakkers and Poisson (1998) found an almost quadratic dependence of
quadrupole moment on the Kerr parameter at small~$j$. In SS~98, the dependence $b(j)$
was studied for normal sequences with $1.4M_{\odot}$ in the static limit and for normal
sequences unstable in the static limit \emph{over the entire range of Kerr parameters}
in which the equatorial rotation velocity on the stellar surface is lower than the
equatorial Keplerian velocity.

\section{NODAL PRECESSION OF ORBITS \\ \protect WITH ARBITRARY INCLINATIONS \\ \protect TO THE EQUATORIAL
PLANE \\ \protect IN THE POST-NEWTONIAN
APPROXIMATION}

Consider the non-Kerr terms in Eqs.~(9) as small Kerr metric perturbations:
$F=F_k(1+v)$; $\Gamma=\Gamma_k(1+z)$; $W=W_k(1+w)$. The eikonal equation (10) in
Boyer--Lindquist coordinates can be written as

\begin{gather}
\hspace{-12pt}(\tilde{S}_{,r}^2\Delta\hspace{-1pt}+\hspace{-1pt}
\tilde{S}_{,\theta}^2)(1\hspace{-1pt}+\hspace{-1pt}v\hspace{-1pt}-\hspace{-1pt}z)\hspace{-2pt}
=\hspace{-2pt}(E(r^2\hspace{-2pt}+\hspace{-1pt}j^2)\hspace{-2pt}-
\hspace{-2pt}L_j)^2/\Delta\hspace{-2pt}-\\
{}-
(L-jE)^2+\cos^2\theta\left((E^2-1)j^2-\frac{L^2}{\sin^2\theta}\right)-\nonumber\\
-v\left((\frac{(r^2+j^2)^2}{\Delta}-j^2\sin^2\theta)E^2+
\frac{L^2}{\sin^2\theta}-\nonumber\right.\\
{}-\left.\frac{L^2j^2}{\Delta}\right)-\frac{4w(2j^2r^2\sin^2\theta+ELjr)}{\Delta}.\nonumber
\end{gather}
The asymptotic result for the nodal precession frequency that follows from Eq.~(22) can
also be obtained as follows.

Consider the energy integral for a particle moving in the gravitational field of a point
mass with a quadrupole moment in the Newtonian approximation:

\begin{gather}
\frac{1}{2}\left(r^2\left(\frac{d\theta}{dt}\right)^2+\frac{L^2}{r^2\sin^2\theta}+
\left(\frac{dr}{dt}\right)^2\right)-\frac{1}{r}\\
{}+\frac{b}{2r^3} (3\cos^2\theta-1)=H,\quad
H=\mathrm{const}.\nonumber
\end{gather}
In the absence of quadrupole moment for circular orbits, we have

\begin{gather}
H=-1/2;\quad p_r=0;\quad L^2=r\sin^2s;\\
 \cos\theta=\cos
s\cos(t/r^{3/2}).\nonumber
\end{gather}
We restrict our analysis to the case $b/r^2<<1$ and denote $\tau\equiv t/r^{3/2}$. The
energy integral, to within quantities of the second order of smallness in $b/r^2$, can
be written as

\begin{gather}
\left(\frac{d\theta}{d\tau}\right)^2=\left(1-\frac{L^2}{r\sin^2\theta}\right)
\left(1-4\frac{\delta r}{r}\right)\\
{}+2r\delta H-\frac{b}{r^2}(3\cos^2\theta-1).\nonumber
\end{gather}
We write the constant~$H$ via its value at the turning point$d\theta/d\tau=0$:

\begin{gather}
1+rH=\frac{L^2}{r\sin^2 s}+\frac{b}{2r^2}(3\cos^2s-1).
\end{gather}
With~(26), the energy integral (25) takes the form

\begin{gather}
\left(\frac{d\theta}{d\tau}\right)^2=
\frac{\sin^2\theta-\sin^2s}{\sin^2\theta}\\
{}\times \left(\frac{L^2}{r\sin^2s}-4\frac{\delta
r}{r}+\frac{3b}{r^2}\sin^2\theta\right).\nonumber
\end{gather}
An analog of circular orbits in a quadrupole field are perturbed orbits with a zero mean
deviation $\delta r$ from a sphere of constant radius:

\begin{gather}
\int\delta r \,d\,\phi(t)=0.
\end{gather}
In Eq.~(28), the integral is taken over the meridional period $T_{\theta}$ in which the
particle again returns to the minimum angle~$s$. Given~(28), the azimuthal angle $\phi$
changes in time $T_{\theta}/4$ by

\begin{gather}
\hspace{-11pt}\Delta\phi=\int_s^{\pi/2}\hspace{-2pt}
\sqrt{\frac{L^2}{r\sin^2s}}\frac{d\theta}{\sin^2\theta}
\left((\cot^2 s-\cot^2\theta)\times\right.\\ \left.
\hspace{-2pt}{}\times\hspace{-2pt}\left(\frac{L^2}{r\sin^2s}+
\frac{3b}{r^2}\sin^2\theta\right)\right)^{1/2}\hspace{-2pt}\approx\frac{\pi}{2}
\hspace{-2pt}\left(1-\frac{3b}{2r^2}\sin
s\right).\nonumber
\end{gather}
Consequently, the nodal precession because of NS oblateness (the presence of a
quadrupole moment) is

\begin{gather*}
2\pi\nu_{\mathrm{nod}}=-\frac{3}{2r^{7/2}}b\sin s.
\end{gather*}
Combining this result with the formula for the precession frequency of inclined circular
orbits in a Kerr field [see Eq.~(52) in Sibgatullin (2001)], we \textbf{derive the
sought-for formula for the nodal precession frequency in the post-Newtonian
approximation}:

\begin{gather}
2\pi\nu_{\mathrm{nod}}=-\frac{2j}{r^3}-\frac{3}{2r^{7/2}}(b+j^2)\sin s.
\end{gather}
Formula~(30) at $s=\pi/2$ transforms to the formula of Markovic (2000). In exactly the
same way, for the azimuthal (Keplerian) period at an inclined orbit in the field of an
oblate star in the post-Newtonian approximation, we obtain using (28)

\begin{gather}
\hspace{-9pt}T/(2\pi) = r^{3/2} + (3\sin{s} -2) j -\f34 \f{b}{r^{1/2}}
(2 -\sin{s}) + \f{9}{4\, r^{1/2}} \cos^2{s}\,(j^2 +b).
\end{gather}
At $s\to\pi/2$, Eq.~(31) matches Eq.~(20) in the principal terms.

\section*{ACKNOWLEDGMENTS}

I wish to thank Prof. Sunyaev for fruitful discussions.

Translated by V. Astakhov


\begin{thebibliography}{99}

\bibitem{8:Sibgatullin_n}
 G.~B.~Cook, S.~L.~Shapiro, and S.~A.~Teukolsky, Astrophys. J.
\textbf{424}, 823 (1995).

\bibitem{30:Sibgatullin_n}
 W.~Cui, S.~N.~Zhang, and W.~Chen, Astrophys. J. Lett. \textbf{492},
L53 (1998);  astro-ph/9811023.

\bibitem{5:Sibgatullin_n}
 A. G. Doroshkevich, Ya. B. Zel'dovich, and I. D. Novikov, Zh. @Eksp.
Teor. Fiz. \textbf{49}, 170 (1965)  [Sov. Phys. JETP \textbf{22}, 122
(1966)].

\bibitem{3:Sibgatullin_n}
 V. L. Ginzburg and L. M. Ozernoi, Zh. Eksp. Teor. Fiz. \textbf{47},
1030 (1964)  [Sov. Phys. JETP \textbf{20}, 689 (1964)].

\bibitem{4:Sibgatullin_n}
 M.~Johnston and R.~Ruffini, Phys. Rev. D \textbf{10}, 2324 (1974).

\bibitem{6:Sibgatullin_n}
 S.~Kato, Publ. Astron. Soc. Jpn. \textbf{42}, 99 (1990).

\bibitem{7:Sibgatulin_n}
 D.~Kramer and G.~Neugebauer, Phys. Lett. A \textbf{75}, 259 (1980).

\bibitem{9:Sibgatulin_n}
 W.~Laarakkers and E.~Poisson, Astrophys. J. \textbf{512}, 282 (1999);
 gr-qc/9709033 (1998).

\bibitem{10:Sibgatulin_n}
 J.~Lense and H.~Thirring, Phys. Z. \textbf{19}, 156 (1918).

\bibitem{11:Sibgatulin_n}
 V.~S.~Man'ko and E.~Ruis, Class. Quantum Grav. \textbf{15}, 2007
(1998).

\bibitem{12:Sibgatulin_n}
 V.~C.~Man'ko, N.~R.~Sibgatullin, \emph{et al.}, Phys. Rev. D
\textbf{49}, 5144 (1994).

\bibitem{13:Sibgatulin_n}
 D.~Markovic, astro-ph/0009450.

\bibitem{14:Sibgatulin_n}
 A.~Merloni, M.~Vietri, L.~Stella, and D.~Bini, Mon. Not. R. Astron.
Soc. \textbf{304}, 155 (1999);  astro-ph/9811198.

\bibitem{15:Sibgatulin_n}
 Ch.~W.~Misner, K.~S.~Thorne, and J.~A.~Wheeler, \emph{Gravitation}
(Freeman, New York, 1973).

\bibitem{16:Sibgatulin_n}
 Sh.~M.~Morsink and L.~Stella, Astrophys. J. \textbf{513}, 827 (1999);
 astro-ph/9808227.

\bibitem{17:Sibgatulin_n}
 A.~T.~Okazaki, S.~Kato, and J. Fukue, Publ. Astron. Soc. Jpn.
\textbf{39}, 457 (1987).

\bibitem{18:Sibgatulin_n}
 D.~Psaltis, R.~Wijnands, J.~Homan, \emph{et al.}, Astrophys. J.
\textbf{520}, 763 (1999);  astro-ph/9903105.

\bibitem{19:Sibgatulin_n}
 F.~D.~Ryan, Phys. Rev. D \textbf{52}, 5707 (1995).

\bibitem{20:Sibgatulin_n}
 F.~D.~Ryan, Phys. Rev. D \textbf{55}, 6081 (1997).

\bibitem{31:Sibgatulin_n}
 N. I. Shakura and R. A. Sunyaev, Astron. Astrophys. \textbf{24}, 337
(1973).

\bibitem{32:Sibgatulin_n}
 N. I. Shakura and R. A. Sunyaev, Mon. Not. R. Astron. Soc.
\textbf{175}, 613 (1976).

\bibitem{33:Sibgatulin_n}
 M.~Shibata and M.~Sasaki, Phys. Rev. D \textbf{58}, 10401 (1998).

\bibitem{21:Sibgatulin_n}
 N. R. Sibgatullin, \emph{Oscillation and Waves in Strong Gravitational
and Electromagnetic Fields} (Springer Verlag, Heisenberg, 1991) .

\bibitem{22:Sibgatulin_n}
 N. R. Sibgatullin, Pis'ma Astron. Zh. \textbf{27}, 918 (2001) 
 [Astron. Lett. \textbf{27}, 799 (2001) ].

\bibitem{23:Sibgatulin_n}
 N. R. Sibgatullin and R. A. Sunyaev, Pis'ma Astron. Zh. \textbf{24},
894 (1998)  [Astron. Lett. \textbf{24}, 774 (1998)].

\bibitem{24:Sibgatulin_n}
 N. R. Sibgatullin and R. A. Sunyaev, Pis'ma Astron. Zh. \textbf{26},
813 (2000a)  [Astron. Lett. \textbf{26}, 699 (2000a)].

\bibitem{25:Sibgatulin_n}
 N. R. Sibgatullin and R. A. Sunyaev, Pis'ma Astron. Zh. \textbf{26},
899 (2000b)  [Astron. Lett. \textbf{26}, 772 (2000b)].

\bibitem{26:Sibgatulin_n}
 L.~Stella, astro-ph/0011395.

\bibitem{27:Sibgatulin_n}
 L.~Stella and M.~Vietri, Astrophys. J. Lett. \textbf{492}, L59 (1998);
 asto-ph/9709085.

\bibitem{28:Sibgatulin_n}
 L.~Stella, M.~Vietri, and Sh.~M.~Morsink, Astrophys. J. \textbf{524},
L63 (1999);  asto-ph/9907346).

\bibitem{29:Sibgatulin_n}
 N.~Stergioulas, http://pauli.phys.uwm.edu/Code/rns;
www.livingreviews.org/Articles/ Volume1/ 1998--8stergio.

\bibitem{1:Sibgatulin_n}
 M.~van~der~Klis, astro-ph/0001167.

\bibitem{2:Sibgatulin_n}
 M.~H.~van~Kerkwijk, Deepto~Chakrabarty, J.~E.~Pringle, and
R.~A.~M.~Wijers, Astrophys. J. Lett. \textbf{499}, L67 (1998); 
astro-ph/9802162.

\end{thebibliography}
\end{document}